\documentclass[twocolumn,floatfix,showpacs,floats,aps,prb,showpacs,amsfonts,amssymb]{revtex4}
\usepackage{graphicx}
\usepackage{hyphenat}
\usepackage{color}
\usepackage{psfrag}
\usepackage[latin1]{inputenc}
\usepackage{amssymb}
\newlength{\sepmod}
\begin{document}

%\begin{frontmatter}

\title{Frustration effects in antiferromagnets on planar random graphs}

\author{Martin Weigel}
%\ead{M.Weigel@ma.hw.ac.uk}
\email{M.Weigel@ma.hw.ac.uk}
%\and
\author{Des Johnston}
%\ead{M.Weigel@ma.hw.ac.uk}
\email{D.A.Johnston@ma.hw.ac.uk}
%\address{Department of Mathematics and the Maxwell Institute for Mathematical
%  Sciences, Heriot-Watt University, Edinburgh, EH14~4AS, UK}
\affiliation{Department of Mathematics and the Maxwell Institute for Mathematical
  Sciences, Heriot-Watt University, Edinburgh, EH14~4AS, UK}

\begin{abstract}
  We consider the effect of geometric frustration induced by the random distribution
  of loop lengths in the ``fat'' graphs of the dynamical triangulations model on
  coupled antiferromagnets. While the influence of such connectivity disorder is rather
  mild for ferromagnets in that an ordered phase persists and only the properties of
  the phase transition are substantially changed in some cases, any
  finite-temperature transition is wiped out due to frustration for some of the
  antiferromagnetic models. A wealth of different phenomena is observed: while for
  the annealed average of quantum gravity some graphs can adapt dynamically to allow
  the emergence of a N\'eel ordered phase, this is not possible for the quenched
  average, where a zero-temperature spin-glass phase appears instead. We relate the
  latter to the behaviour of conventional spin-glass models coupled to random graphs.
\end{abstract}

%\begin{keyword}
%  Ising model \sep geometric frustration \sep quantum gravity \sep
%  random graphs 
%  \PACS 04.60.Nc \sep 75.10.Hk \sep 75.50.Lk
%04.60.Nc Lattice and discrete methods
%75.10.Hk Classical spin models
%05.10.Ln Monte Carlo methods  
%68.35.Rh Phase transitions and critical phenomena
%75.50.Lk Spin glasses and other random magnets
%64.60.Ak Percolation (phase transitions)  
%\end{keyword}


\maketitle

%\end{frontmatter}

\section{Introduction}

Models of random networks and surfaces have received extensive attention in
statistical physics and field theory due to their wealth of applications in such
diverse fields as the modelling of the internet \cite{albert:02a}, spin glass physics
\cite{mezard:03} and quantum gravity \cite{ambjorn:book}. The diversity of
applications is reflected in a rather large variety of different graph models
considered. Generic random graph and network models include the most general
Erd\"os-R\'enyi model of $n$ bonds distributed randomly between pairs selected from
$N$ vertices, as well as the Barab\'asi-Albert graphs constructed with the
preferential attachment rule or the Watts-Strogatz method of interpolating between a
regular lattice and a random graph via rewiring of links \cite{albert:02a}. These
constructions result in variable, but {\em uncorrelated\/} vertex degrees with graph
ensembles fully defined by the co-ordination number distribution $P(q)$. On the other
hand, fixed-degree or $k$-regular random graphs have also been considered, which are
equivalent to the non-planar or ``thin'' $\phi^3$, $\phi^4$, $\ldots$ Feynman
diagrams of a zero-dimensional field theory \cite{bachas:94}. While none of these
networks feature a well-defined topology allowing for a local geometrical
interpretation, a fattening of the thin graph propagators to ribbons yields
orientable faces, enabling an interpretation of the graphs as random surfaces of
fixed genus. The resulting dynamical triangulations model has been thoroughly studied
with matrix model and combinatorial methods in the context of Euclidean quantum
gravity \cite{ambjorn:book}. The topological constraint induces spatial correlations
in the degree distribution, and the ensemble of graphs is no longer fully determined
by $P(q)$. Other ensembles of random graphs of well-defined topology have been
studied, for instance the Poissonian Vorono\"{\i}-Delaunay tessellations resulting
from a generalisation of the crystallographic Wigner-Seitz construction to randomly
distributed generators \cite{okabe:book}.

A lot of effort has been invested in understanding the generic geometrical properties
of these graph ensembles. In particular, for the network-type models dynamical
properties have been regularly considered, such as the emergence of a giant component
as edges are successively added to the graphs \cite{albert:02a}. For the Delaunay and
dynamical tessellations with a well-defined topology, genuinely geometrical
attributes such as the fractal or Hausdorff dimension and the correlation function of
the local degrees have been of most interest \cite{ambjorn:book,okabe:book}. As an
alternative to direct investigations, these ensembles might be characterised by
observing the cooperative behaviour of matter variables such as classical Ising spins
$s_i = \pm 1$ with Hamiltonian
\begin{equation}
  \label{eq:ising_hamiltonian}
  {\mathcal H} = -\sum_{\langle i,j\rangle}J_{ij} s_i s_j,
\end{equation}
or more general Potts or O($n$) spins placed on the graph vertices, assuming
ferromagnetic couplings $J_{ij} = J_0 > 0$. This exercise is also of interest from
the inverted point-of-view of studying the effect of these various types of
connectivity disorder on the spin models, being complementary to the canonical case
of weak disorder from randomness of the couplings on regular lattices
\cite{harris:74a}.  Graphs without topological constraint are inherently non-local
and thus effectively infinite-dimensional. Consequently, one expects mean-field phase
transitions for the ferromagnetic models. These are indeed found for the
Erd\"os-R\'enyi and $k$-regular (or thin) random graphs
\cite{dominicis:89,bachas:94,baillie:94,johnston:97a,herrero:02,mezard:03},
transforming them into convenient alternatives to treatments on the Bethe lattice or
the complete graph, not encumbered with boundary effects. More generally, degree
distributions with divergent moments can lead to interesting deviations from
mean-field behaviour \cite{goltsev:03}. The effects on surface-like graphs are less
homogeneous. For (uncorrelated) bond disorder, the celebrated Harris criterion
\cite{harris:74a} predicts a change of critical behaviour for cases with a positive
specific-heat exponent $\alpha$. An analogous criterion can be formulated for random
graphs, taking into account the spatial correlation of the degree distribution
\cite{wj:04a}. The predicted change in universality class for virtually all types of
matter coupled to dynamical triangulations or fat graphs is in agreement with exact
results for percolation \cite{kazakov:89a} and numerical investigations of the Ising
and Potts models \cite{wj:00a}. For Vorono\"{\i}-Delaunay tessellations, however, the
predicted change of universality class for models with $\alpha > 0$ such as the
two-dimensional $q = 3$, $4$ Potts and three-dimensional Ising models is not observed
numerically \cite{wj:02b,wj:03a}. It is an open problem why, instead, these models
behave according to their regular lattice critical exponents.

In contrast to this weak-disorder case, the effect of connectivity disorder on models
with {\em antiferromagnetic\/} interactions is much more profound. The existence of
odd-length loops on many of the graphs discussed leads to severe geometric
frustration with the possibility of altogether precluding the onset of a long-range
ordered phase. Interestingly, however, this problem for random graphs has received
very little attention to date \cite{baillie:94}, such that the behaviour of, e.g.,
the Ising antiferromagnet for the various cases is unclear. The effect of tuning the
amount of frustration on a {\em regular\/} lattice has been considered in
Ref.~\onlinecite{poulter:01} for the $\pm J$ Ising model on the triangular lattice,
where it was found that a spin-glass phase appears as the concentration $p$ of
antiferromagnetic bonds is increased from zero and disappears again as the system
comes close to the perfect frustration of the pure antiferromagnet. Some use has been
made of $k$-regular or thin random graphs as an alternative realisation of the
mean-field limit of spin glasses \cite{dominicis:89,mezard:03}, and in general one
expects the perfect frustration of the effectively infinite-dimensional graphs of the
non-topological type to lead to mean-field spin-glass behaviour for the
antiferromagnet. The less extreme case of fixed-topology graphs is the subject of the
present study.

The rest of the paper is organised as follows. Section \ref{sec:frust} discusses the
general problem of frustration exerted by graphs of the dynamical triangulations type
on antiferromagnets and the annealed and quenched limits for the Ising
antiferromagnet coupled to different graph types. The results of Monte Carlo
simulations for the annealed case are discussed. In Sec.\ \ref{sec:zero}, we
investigate the quenched limit by means of exact ground-state computations in the
framework of a defect-wall calculation, comparing Gaussian and bimodal spin glasses
to the antiferromagnet. Finally, Sec.\ \ref{sec:concl} contains our conclusions.

\section{Frustration from fat graphs\label{sec:frust}}

\begin{figure}[tb]
  \centering
  \begin{tabular}{rrr}
    \includegraphics[clip=true,keepaspectratio=true,width=2.5cm]{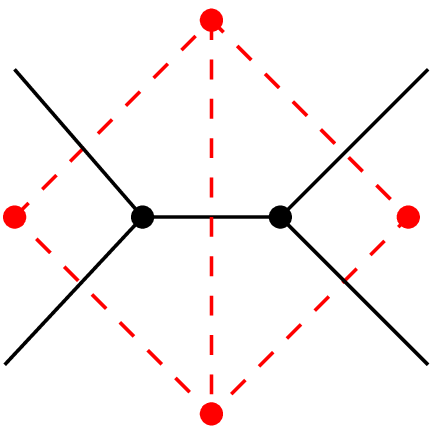} &
    \raisebox{1.12cm}{\hspace{0.25cm}\includegraphics[clip=true,keepaspectratio=true,width=1.25cm]{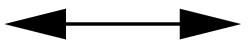}} &
    \includegraphics[clip=true,keepaspectratio=true,width=2.5cm]{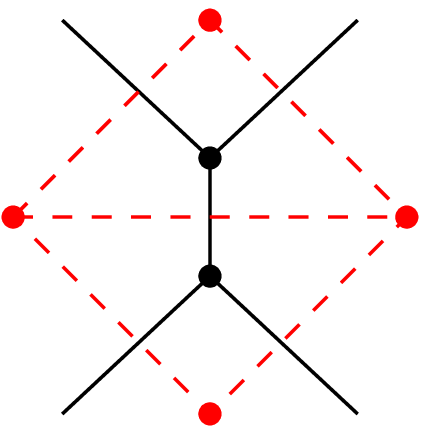}
  \end{tabular}
  \vspace*{-0.4cm}
  \caption
  {The link-flip move on two adjacent triangles of a dynamical triangulation (dashed
    lines). The solid lines denote the corresponding dual, three-valent, ``fat''
    $\phi^3$ graphs.}
  \label{fig:phi3_move}
\end{figure}

\begin{figure}[tb]
  \centering
  \includegraphics[clip=true,keepaspectratio=true,width=7cm]{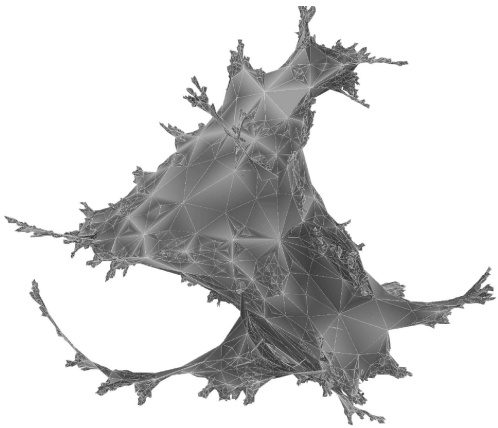}
  \caption
  {Example of a dynamical triangulation of $N_2 = 5\,000$ triangles embedded into
    Euclidean space. Note that the equilaterality requirement for the triangles is
    only approximately fulfilled since an exact embedding into $\mathbb{R}^3$ is not
    guaranteed to exist.\label{fig:snapshot}}
\end{figure}

Dynamical triangulations originate in the discrete approach to Euclidean quantum
gravity via a path-integral quantisation of the gravitational interaction. The
integral over metrics is regularised by a sum over combinatorial manifolds or
simplicial complexes \cite{ambjorn:book}. We focus here on the case of two
dimensions, where this amounts to a sum over all possible gluings of equilateral
triangles to closed, orientable surfaces of a given, for instance planar or toroidal,
topology. Graphs of this type are hence constructed by taking a number $N_2$ of
equilateral triangles and connecting them together along their edges at random, such
that the resulting graph represents a closed and orientable surface of well-defined
topology. In a computer simulation, such graphs can be implemented by the successive
application of the ``link-flip'' move depicted in Fig.~\ref{fig:phi3_move} to the
graph starting, e.g., from a regular hexagonal lattice with periodic boundary
conditions. The geometric duals of these triangulations naturally are graphs of fixed
degree $k=3$, known as planar $\phi^3$ Feynman diagrams in field theory. The
planarity constraint necessitates a ``fattening'' of the usual propagators to
ribbons, such that this type of diagrams is often refered to as ``fat graphs''
\cite{zinn-justin}. Instead of triangles, one can, of course, consider more general
elementary polygons, such as the quadrangulations constructed from gluings of
squares, whose duals are then consequently ``fat'' $\phi^4$ graphs etc. The resulting
combinatorial problem of counting these discrete surfaces can be solved exactly for
various cases using a matrix-model formulation \cite{brezin:78a} or alternative
combinatorial approaches \cite{tutte:62a,bouttier:02}. Geometrically, the most
striking result is that of an unusually large internal Hausdorff dimension $d_h = 4$
for these topologically two-dimensional surfaces \cite{ambjorn:book}, resulting from
a structure of ``baby universes'' connected to the main graph body with a minimal
number of links (``bottlenecks''). This fractal structure is apparent from the
example graph embedding shown in Fig.\ \ref{fig:snapshot}: while the graphs
themselves are, by construction, defined in a purely intrinsic manner, without
reference to an embedding into external space, an approximate embedding to some
extent observing the constraint of equal edge lengths can be constructed to visualise
the internal geometry \cite{doktor}. The distribution $P(q)$ of co-ordination numbers
is found to fall off exponentially \cite{boulatov:86a,doktor} and hence does not
exhibit the fat tails found in small-world networks. However, the structure of the
graphs is not solely determined by $P(q)$, since the well-defined topology introduces
long-range correlations in the co-ordination numbers, declining as $r^{-2}$ at large
separations $r$.  \cite{wj:04a}

\subsection{Annealed average}

When coupling matter variables to the graphs, averages have to be performed for the
fluctuations of the graphs as well as those of the coupled spins with associated
characteristic time scales $\tau_\mathrm{graph}$ and $\tau_\mathrm{matter}$. The {\em
  annealed\/} case of $\tau_\mathrm{graph} = \tau_\mathrm{matter}$ is the situation
considered in the context of matter coupled to quantum gravity.  Since spin and
disorder degrees-of-freedom fluctuate on the same time scales, the total free energy
of the system is given by
\begin{equation}
  \label{eq:annealed}
  F_\mathrm{annealed}(\beta) = -\beta^{-1} \ln\langle Z(\beta, {\cal G})\rangle_{\cal G},
\end{equation}
where $Z(\beta, {\cal G})$ denotes the partition function for spins on a fixed graph
${\cal G}$ and $\langle\cdot\rangle_{\cal G}$ symbolises the average over the
considered graph ensemble.  Hence, the effect of geometric randomness on the
behaviour of spins is augmented by a {\em back-reaction\/} of matter variables on the
underlying graphs.  Formulating the combined problem as a multi-matrix model, several
cases such as the (ferromagnetic) Ising \cite{kazakov:86a,burda:89a}, $q$-state
Potts- \cite{daul:95a,zinn-justin:00b} and O($n$) models \cite{gaudin:89a,eynard:95a}
(where the two latter classes, of course, include the simpler Ising model as the
special cases $q=2$ resp.\ $n=1$) can be treated analytically (in the regime where
they show continuous phase transitions on regular lattices, i.e., for $q\le 4$ resp.\
$-2\le n\le 2$). In these cases, the coupling to dynamical triangulations results in
a shift of universality class. More generally, in the framework of Liouville theory,
the dressing of conformal weights $\Delta$ of critical matter coupled to
two-dimensional quantum gravity is predicted to be
\cite{knizhnik:88a,david:88a,distler:89a}
\begin{equation}
  \tilde{\Delta} = \frac{\sqrt{1-c+24\Delta}-\sqrt{1-c}}{\sqrt{25-c}-\sqrt{1-c}},
  \label{dressing_KPZ}
\end{equation}
where $c$ denotes the central charge \cite{henkel:book}. The resulting critical
exponents agree with the exact results discussed above, and the predictions for a
number of cases have been confirmed by numerical simulation studies, see, e.g.,
Refs.~\onlinecite{jurkiewicz:88a,baillie:92c,catterall:93a}.

Coupling {\em antiferromagnets\/} to this type of random graphs leads to strong
frustration \cite{toulouse:77a} due to the presence of loops of odd length. Depending
on the exact type of graphs considered, however, the annealed average to some extent
allows for the geometry to adapt to the antiferromagnetic interactions. In
particular, consider an Ising antiferromagnet according to
Eq.~(\ref{eq:ising_hamiltonian}) with $J_{ij} = -J_0$ for all bonds, coupled to the
following dynamical graphs:

\begin{itemize}
\setlength{\itemsep}{1.5ex}
\item {\bf Triangulations}: Here, all elementary faces of the graph are frustrated,
  and the frustration cannot be relieved by a dynamic response of the lattice. As has
  been shown by Wannier \cite{wannier:50}, the Ising antiferromagnet on a {\em
    regular\/} triangular lattice remains paramagnetic down to zero temperature
  where, due to frustration, the system has a finite residual entropy. It is clear
  that for any planar triangulation (whether regular or random) configurations of
  minimal energy have two satisfied and one broken bond in each triangle. A large
  number of such states exists already for the triangular lattice, which is a member
  of the ensemble of dynamical triangulations considered. The freedom of changing
  spin configurations without leaving the ground-state manifold is found to be local,
  offering no energetic reward and thus precluding long-range order. In fact, the
  residual, zero-temperature entropy of this model has been calculated
  \cite{bachas:91a} to be $S_0 \approx 0.2613$, to be compared to the value $S_0
  \approx 0.3383$ found for the triangular antiferromagnet \cite{wannier:50}.
\item {\bf Quadrangulations}: Any planar quadrangulation is bipartite --- it cannot
  contain any loops of odd length since all loops are composed of the elementary
  faces of length four. This structure allows for a two-colouring of the lattice
  vertices in, say, black and white sites. Introducing extra (non-fluctuating) signs
  $\epsilon_i = \pm 1$ at the vertices and performing a Mattis transformation as
  \cite{mattis:76}
  \begin{equation}
    \label{eq:mattis}
    s_i' = \epsilon_i s_i,\;\;\;J_{ij}' = J_{ij}\epsilon_i\epsilon_j,
  \end{equation}
  leaves the Hamiltonian (\ref{eq:ising_hamiltonian}) and thus the partition function
  invariant. Choosing $\epsilon_i^W = -1$ for the white and $\epsilon_i^B = +1$ for
  the black vertices maps the system identically to the Ising ferromagnet with
  $J_{ij} = +J_0$. Hence, for the random lattice the results of
  Refs.~\onlinecite{kazakov:86a,burda:89a} apply, and the system undergoes a third-order
  phase transition to a N\'eel ordered state.
\item {\bf $\phi^3$ or $\phi^4$ graphs}: While neither of these graph classes is
  bipartite as a whole, it is clear that, e.g., with the hexagonal lattice ($\phi^3$
  graphs) resp.\ the square lattice ($\phi^4$ graphs) bipartite graphs occur in
  both ensembles. Hence the ground states of either system are perfectly N\'eel
  ordered configurations, and it is reasonable to expect long-range order to persist
  for some finite range of temperatures.
\end{itemize}

\begin{figure}[tb]
  \centering
  \includegraphics[clip=true,keepaspectratio=true,width=7cm]{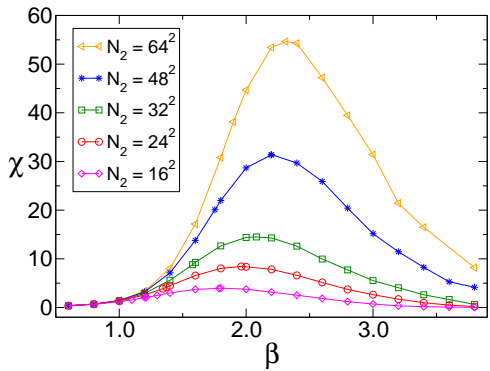}
  \caption
  {Magnetic susceptibility $\chi = N_2(\langle m^2\rangle-\langle |m|\rangle^2)$ of
    the Ising antiferromagnet coupled to annealed fat $\phi^3$ graphs with $N_2 =
    16^2$ up to $N_2 = 64^2$ vertices.}
  \label{fig:annealed_susc}
\end{figure}

\begin{figure}[tb]
  \centering
  \includegraphics[clip=true,keepaspectratio=true,width=7cm]{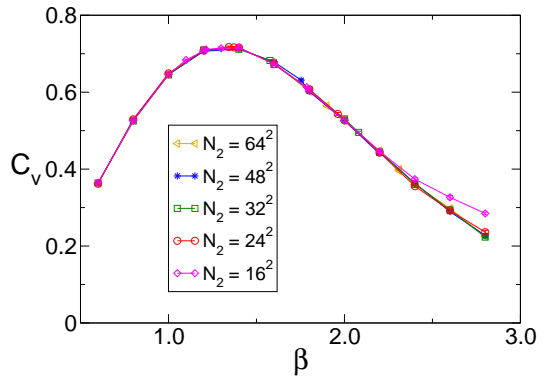}
  \caption
  {Specific heat $C_v = N_2\beta^2(\langle e ^2 \rangle - \langle e\rangle^2)$ of the
    Ising antiferromagnet coupled to annealed fat $\phi^3$ graphs from Monte Carlo
    simulations.}
  \label{fig:Cv}
\end{figure}

To investigate this last case of a non-trivial interaction between geometric
frustration and antiferromagnetism, we performed Monte Carlo simulations of the
coupled system for graph sizes of $N_2 = 16^2$ up to $128^2$ vertices. As the
temperature is lowered, we indeed find the emergence of antiferromagnetic N\'eel
order with signatures of a phase transition at rather low temperatures $\beta_c
\approx 2.5$, cf.\ the susceptibility data shown in Fig.~\ref{fig:annealed_susc}.
This temperature is to be compared with the phase transition at $\beta_c =
\frac{1}{2}\ln\frac{108}{23} \approx 0.773$ found for the Ising {\em ferromagnet\/}
on the ``fat'' $\phi^3$ graphs without self-energy and tadpole insertions considered
here \cite{burda:89a}. We find the specific heat to be completely independent of
system size with a broad peak around $\beta \approx 1.4$, cf.\ Fig.\ \ref{fig:Cv},
and no crossing of the Binder parameter within the temperature range $0.1\le \beta
\le 4.0$ considered.  These findings, together with the apparent scaling of the
susceptibility on the low-temperature side of the peak seen in
Fig.~\ref{fig:annealed_susc}, hint at the presence of a critical (low-temperature)
phase with an associated Kosterlitz-Thouless phase transition \cite{weigel:04b}. This
conjecture is further supported by the very slow convergence of effective critical
temperatures as the system size is increased, compatible with a logarithmic rather
than a power-law approach. A series of simulations for the {\em ferromagnet\/} on the
same lattices performed for comparison, on the other hand, yields $\beta_c =
0.783(7)$, $d_h \nu = 3.3(2)$ and $\gamma/d_h \nu = 0.68(3)$, perfectly compatible
with the exact results \cite{kazakov:86a,burda:89a}.

\begin{figure}[tb]
  \centering
  \includegraphics[clip=true,keepaspectratio=true,width=7cm]{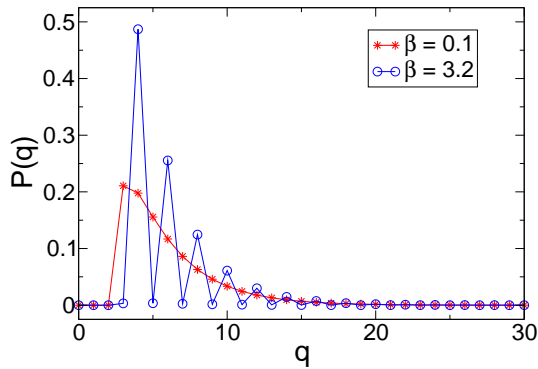}
  \caption
  {Distribution $P(q)$ of loop lengths of annealed, fat $\phi^3$ random graphs
    coupled to an Ising antiferromagnet in the high-temperature phase at $\beta =
    0.1$ and deep in the ordered phase at $\beta = 3.2$. Data for graphs with $N_2 =
    32^2$ vertices.}
  \label{fig:annealed_coord}
\end{figure}

The mechanism of the antiferromagnetic transition can be understood from an
inspection of the distribution $P(q)$ of loop lengths: as the temperature is lowered,
the Ising antiferromagnet forces all odd-length loops out of the system, leading to
surfaces completely composed out of faces of even length, i.e., squares, hexagons,
octagons etc., cf.\ Fig.~\ref{fig:annealed_coord}. Thus, the back-reaction of the
antiferromagnetic matter on the graphs drives them into a non-frustrating phase of
bipartite graphs compatible with N\'eel order. The resulting strong coupling between
graphs and matter throughout the whole low-temperature region could plausibly give
rise to a critical phase with associated infinite-order phase transition as implied
by the simulation results discussed above. This is in contrast to the Ising
ferromagnet, where strong interaction between matter and geometry is confined to the
vicinity of the critical point.

\subsection{Quenched average}

The (graph-)quenched limit $\tau_\mathrm{graph} \gg \tau_\mathrm{matter}$ of averages
describes the case usually encountered in condensed matter physics, where impurities
or lattice defects are fixed properties of the sample.  Consequently, the average
over disorder should be performed at the level of observable quantities, such as the
free energy and its derivatives, leading to
\begin{equation}
  \label{eq:quenched}
  F_\mathrm{quenched}(\beta) = -\beta^{-1} \langle \ln Z(\beta, {\cal G})\rangle_{\cal G},
\end{equation}
This interchange of logarithm and disorder average as compared to the annealed limit
of Eq.~(\ref{eq:annealed}) often leads to dramatically different properties.
Unfortunately, no exact prediction for ferromagnets in the spirit of
(\ref{dressing_KPZ}) is available here.  Since it is believed that ferromagnets with
weak quenched disorder in two dimensions in general are related to conformal field
theories with central charge $c=0$, an approximation starting from the KPZ/DDK
framework could be derived from Eq.~(\ref{dressing_KPZ}) by setting $c=0$ there
\cite{gurarie:02}, i.e.,
\begin{equation}
  \tilde{\Delta} = \frac{1}{4}(\sqrt{1+24\Delta}-1).
  \label{KPZ_quenched}
\end{equation}
This form can also be more directly motivated by noting that the central charge of
$n$ replicas of matter variables is additive, such that the limit $n\rightarrow 0$ of
the replica trick naturally leads to a central charge $c=0$ in (\ref{dressing_KPZ})
\cite{johnston:92,baillie:94b,wj:99a}. These predictions cannot be confronted with
any exact solutions. Monte Carlo simulations for the Ising and Potts ferromagnets on
quenched dynamical triangulations \cite{wj:00a,wj:00b} show a change in universality
class, in agreement with the adapted relevance criterion for connectivity disorder
discussed above \cite{wj:04a}. The critical exponents, however, although changed from
their regular-lattice values, in general seem not to be correctly predicted by the
form (\ref{KPZ_quenched}).

For the antiferromagnets, frustration from quenched, fat graphs is potentially
stronger than in the annealed case: the frozen-in disorder of the graphs cannot adapt
to alleviate the energy cost of odd-length loops. As before, the degree of
frustration exerted on the Ising antiferromagnet depends on the ensemble of graphs
considered:

\begin{itemize}
\setlength{\itemsep}{1.5ex}
\item {\bf Triangulations}: Following the argumentation presented above, the system
  is paramagnetic at all temperatures for each fixed random triangulation.
  Consequently, the quenched average merely describes the modulation of a completely
  disordered system.
\item {\bf Quadrangulations}: The perfect bipartiteness of the graphs is not affected
  by the process of the disorder average. Via the described Mattis transformation,
  the system is identical to the Ising ferromagnet on quenched quadrangulations. By
  universality, the critical behaviour is expected to be that of the Ising
  ferromagnet on quenched fat $\phi^3$ graphs studied in Ref.~\onlinecite{wj:00a}.
\item {\bf $\phi^3$ or $\phi^4$ graphs}: This graph ensemble features a broad
  distribution of odd-length loops exerting strong frustration on the coupled
  antiferromagnet. The lack of a back-reaction of the Ising spins on the underlying
  graphs prevents the suppression of loops of odd length observed in the annealed
  limit. The bipartite graph configurations selected in the low-temperature phase of
  the latter only occur with vanishing weight in the quenched average. This appears
  to preclude the emergence of long-range order at finite temperatures.  However, the
  frozen-in frustration might give rise to {\em spin-glass order\/} at zero
  temperature.
\end{itemize}

The rest of this paper is devoted to an investigation of the possibility of such
spin-glass order for the Ising antiferromagnet on quenched fat $\phi^3$ graphs.

\section{Zero-temperature phase on quenched fat graphs\label{sec:zero}}

Due to the random distribution of faces of odd (frustrated) and even (unfrustrated)
lengths, it is natural to suspect that the Ising antiferromagnet on $\phi^3$ fat
graphs behaves similarly to a $\pm J$ Edwards-Anderson spin glass on the same
lattices.  It is not {\em a priori\/} clear, however, whether the long-range
correlation of co-ordination numbers \cite{wj:04a} might cause any non-universal
differences between these cases. How can the appearance of spin glass be detected and
its properties determined? Generalising Peierls' argument for the stability of an
ordered phase, a droplet-scaling theory for the spin-glass phase can be formulated
\cite{bray:87a}.  The role of the droplet surface (free) energy is then taken on by
the {\em width\/} $J(L)$ of the distribution of random couplings for a real-space
renormalisation group decimation at length scale $L$. In the course of
renormalisation, $J(L)$ scales as $J(L) \sim L^{\theta_s}$, defining the {\em spin
  stiffness exponent\/} $\theta_s$. If the system scales to weak coupling, $\theta_s
< 0$, spin-glass order is unstable at finite temperature and the system is below its
lower critical dimension, with $\theta_s$ describing the properties of the {\em
  critical\/} point at temperature $T=0$. On the other hand, $\theta_s > 0$ indicates
stability of a spin-glass phase at finite temperature. Numerically, the domain-wall
free energy can be determined from the energy difference between ground states of
systems with different types of boundary conditions (BCs) chosen such as to induce a
relative domain wall \cite{bray:87a}.

\subsection{Method and droplet length scale}

Following this programme, one requires the generation of a number of statistically
independent graph realisations for performing the quenched average. Subsequently, for
each realisation ground states of the Ising antiferromagnet should be calculated. An
ergodic set of graph updates for dynamical triangulations (and the dual $\phi^3$
graphs) is given by the so-called Pachner moves \cite{pachner:91a,doktor}, which we
employ to generate independent realisations of toroidal topology, starting out from a
perfect hexagonal lattice with periodic boundary conditions. Such a toroidal shape is
needed to induce excited states by a change of boundary conditions as indicated
above. The Monte Carlo equilibration is done for closed graphs, leaving the task of
identifying appropriate loops along which to cut them open. Several possibilities
come to mind:

\begin{itemize}
\item In the original hexagonal lattice, the identification of boundaries is obvious.
  If, by construction, no link-flip updates are performed on the links crossing a
  boundary, the toroidal cuts are {\em fixed\/} in the equilibrated graphs. This
  simplicity, however, comes at the expense of strong boundary effects: due to the
  large fractal dimension \cite{ambjorn:book} $d_h =4$, each vertex is
  comparatively closer to one of the boundaries than would be expected for a
  two-dimensional system.
\item Allowing flips of links crossing a boundary, one might keep track of the
  induced evolution of the boundary loops. This turns out to be rather involved, in
  particular due to the appearance of self-intersections of the meandering
  boundary lines. Consequently, we have not considered this approach for the final
  ground-state computations.
\item Ideally, the graph generation should employ periodic boundary conditions
  preventing surface effects with the graphs being ``cut open'' once equilibrated. This
  amounts to the identification of topologically inequivalent elementary loops on the
  surface. It is not immediately obvious, however, how such an optimal {\em
    homotopy\/} basis could be computed for an intrinsic graph structure.
\end{itemize}

\begin{figure}[tb]
  \centering
  \includegraphics[clip=true,keepaspectratio=true,width=6cm]{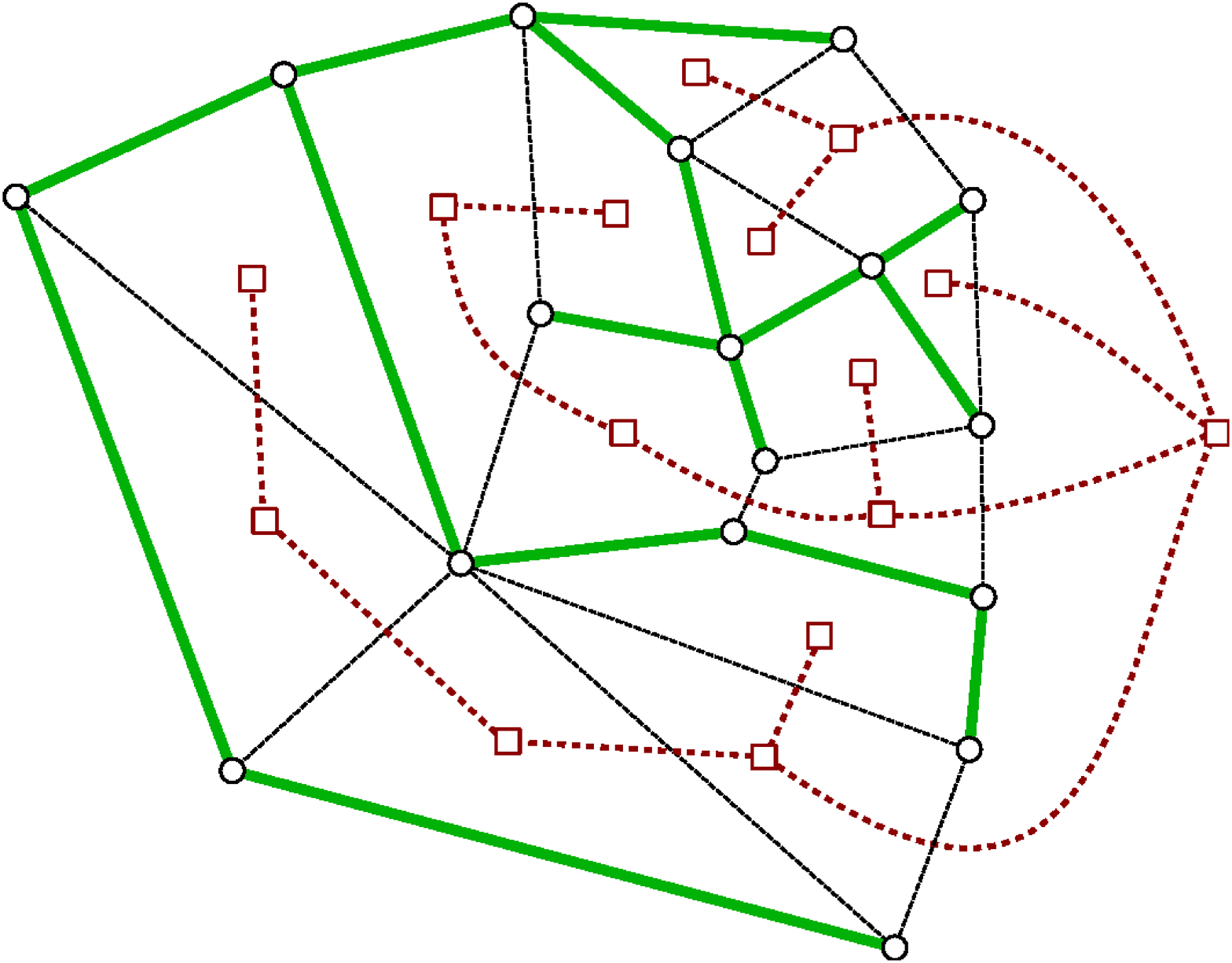}
  \caption
  {On a planar graph, two interdigitating spanning trees on the original and dual
    lattices touch or cross all links.\label{fig:trees}}
\end{figure}

As it turns out, a number of computational problems related to topologically
motivated cuttings of surfaces are {\em NP\/}-hard. In particular, identification of
the so-called ``cut graph'' whose removal renders a surface topologically trivial,
i.e., the locus of all points with at least two shortest paths from a given base
point, is a non-polynomial problem \cite{erickson:04}. Rather surprisingly, however,
a homotopy basis of minimum length in form of a system of loops joint in a common
base point {\em can\/} be computed in polynomial time \cite{erickson:05}. Such a
system consists of $2g$ simple loops (where $g$ is the genus of the surface), whose
complement in the manifold is a topological disk. This decomposition proceeds with a
variant of the tree-cotree decomposition proposed in Ref.~\onlinecite{eppstein:03}:
starting from a marked base point on a planar surface, the dual complement of a
spanning tree is a spanning tree as well, cf.\ Fig.~\ref{fig:trees}. For non-trivial
topology, however, by Euler's formula both trees simply do not have enough edges to
cover the whole graph, and it is easily seen that the edges not touched by either
tree are those defining the topologically non-trivial loops with respect to the base
point. A set of {\em minimal\/} loops can then be found by computing the tree $T$ of
shortest paths in the graph $G$ and the maximum spanning tree $T^\ast$ in the
complement $(G\setminus T)^\ast$, where the weight of each dual edge $e^\ast$ is
chosen to be the length of the shortest loop in $G$ containing the base point as well
as the edge $e$. The desired basis is then given \cite{erickson:05} by the minimal
loops defined by all edges $e$ neither contained in $T$ nor crossed by $T^\ast$.

\begin{figure}[tb]
  \centering
  \includegraphics[clip=true,keepaspectratio=true,width=7cm]{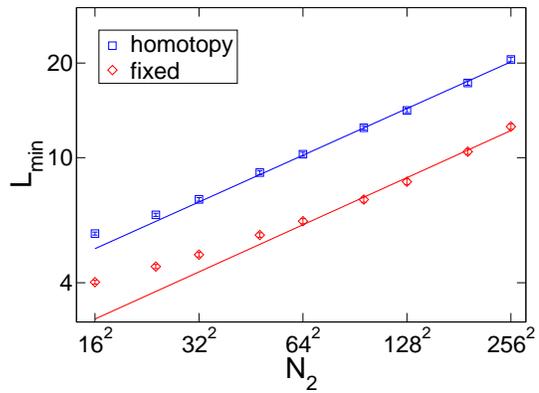}
  \caption
  {Scaling of the average length $\langle L_\mathrm{min}\rangle$ of the domain wall in the
    {\em ferromagnet\/} with antiperiodic boundary conditions together with fits of the form
    $\langle L_\mathrm{min}\rangle = A N_2^{1/d_h'}$. Note the considerably larger scaling
    corrections for the graphs with fixed boundaries.
    \label{fig:hausdorff}}
\end{figure}

Exact ground states of the Ising system on the resulting graphs are computed in
polynomial time via the mapping to a so-called minimum-weight perfect matching
problem \cite{bieche:80a}. Due to a limitation of this approach only topologically
planar graphs can be treated. Hence, after cutting the graph one of the resulting
boundaries is left open, while either periodic (P) or antiperiodic (AP) conditions
are employed along the second boundary. The scaling of the energy difference,
$\langle |\Delta E|_\mathrm{P/AP}\rangle \sim L_\mathrm{DW}^{\theta_s}$, then gives
access to the spin-stiffness exponent $\theta_s$. In contrast to the case of a
regular lattice, where the relevant domain-wall length scale is simply equivalent to
the length of the boundary, $L_\mathrm{DW} \sim L$, the corresponding scale is not
immediately obvious for the fractal graphs at hand. It can be found, however, from
the ground-state problem of the {\em ferromagnet\/}: if AP boundary conditions are
applied along one direction, the system responds with a domain wall of minimum length
$L_\mathrm{min}$, corresponding to the length scale of the applied perturbation. The
corresponding average length $\langle L_\mathrm{min}\rangle$ is shown in
Fig.~\ref{fig:hausdorff} for $5\,000$ graph replicas with $N_2 \le 256^2$.  Here,
$\langle\cdot\rangle$ denotes an average over disorder.  Fits of the form $\langle
L_\mathrm{min}\rangle = A_L N_2^{1/d_h'}$ yield $d_h' = 4.0161(91)$ for the {\em
  fixed\/} boundaries resp.\ $d_h' = 4.0513(72)$ for the {\em homotopy\/} basis,
implying that in fact $d_h' = d_h = 4$, and the relevant domain-wall length scale is
equivalent to the intrinsic length $L_\mathrm{eff} \sim N_2^{1/d_h}$ of the fractal
graphs, implying $\langle |\Delta E|_\mathrm{P/AP}\rangle \sim N_2^{\theta_s/d_h}$.
Note that the applied fits without correction terms allow an impressively precise
determination of $d_h$, while it turned out to be very tedious to determine the
Hausdorff dimension numerically by considering the geometrical two-point function
\cite{ambjorn:95e,doktor}.

\subsection{Spin-stiffness exponent}

\begin{figure}[tb]
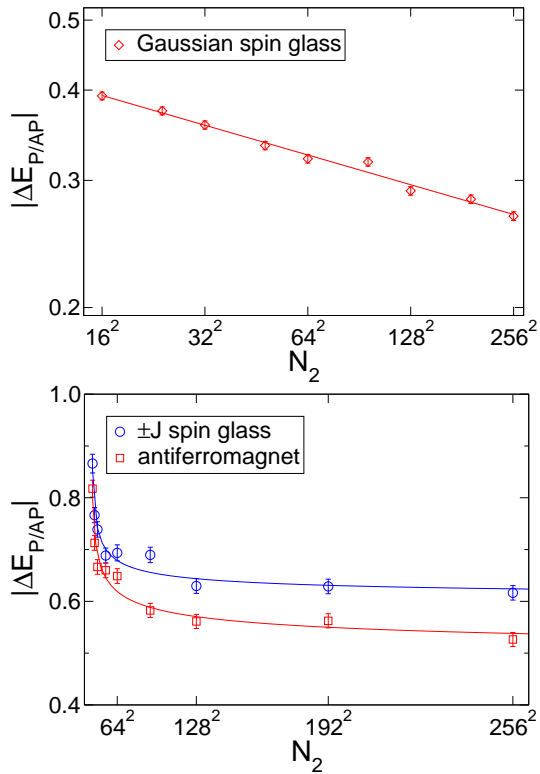

  \centering
  \includegraphics[clip=true,keepaspectratio=true,width=7cm]{homotopy_gauss}\\
  \includegraphics[clip=true,keepaspectratio=true,width=7cm]{homotopy_anti_pmJ}
  \caption
  {Domain-wall energies of the Gaussian and $\pm J$ Ising spin glasses as well as the
    antiferromagnet on quenched, fat $\phi^3$ random graphs. The lines show fits of
    the form $\langle |\Delta E|_\mathrm{P/AP}\rangle = E_0 + S_E N_2^{\theta_s/d_h}$ to the
    data.
    \label{fig:stiffness}}
\end{figure}

To determine the spin-stiffness exponent $\theta_s$, ground-state computations were
performed for a series of graph sizes, comparing the ground-state energies for P and
AP boundaries, using $5\,000$ disorder replica for each size. Since previous work has
almost exclusively focused on the square lattice \cite{kawashima:03a}, as a benchmark
and universality test we first considered the case of the regular honeycomb lattice,
performing fits of the form $\langle |\Delta E|_\mathrm{P/AP}\rangle = E_0 + S_E
L^{\theta_s}$ for $L = 16$, $24$, $\ldots$, $256$. For a Gaussian distribution of the
couplings $J_{ij}$ of (\ref{eq:ising_hamiltonian}), we arrive at $E_0 = 0.0026(261)$,
clearly indicating that asymptotically $E_0 = 0$ as expected. Fixing $E_0 = 0$ then
yields $\theta_s = -0.2843(44)$ with quality-of-fit $Q = 0.29$. This is in excellent
agreement with results for the square lattice \cite{kawashima:03a}, the negative
value of $\theta_s$ indicating that spin-glass order is confined to zero temperature
in this system. For symmetric, bimodal couplings $J_{ij} = \pm J$, on the other hand,
a clear saturation of defect energies is observed, resulting in $E_0 = 0.991(15)$,
leading to a vanishing $\theta_s = 0$ of the effective spin-stiffness exponent. This
is again in agreement with the square-lattice result, showing that the $\pm J$ spin
glass is marginal in two dimensions (but it turns out that, in fact, $T_g = 0$ there)
\cite{kawashima:03a}, see also Refs.~\onlinecite{katzgraber:05a,jorg:06,fisch:06b}.

Although the quenched approximation (\ref{KPZ_quenched}) to the KPZ formula
(\ref{dressing_KPZ}) has been suggested for unfrustrated situations, it is
interesting to see what it predicts for the case at hand. Noting that $\theta_s/d_h =
- 1/d_h \nu = \Delta_\epsilon - 1$, a dressing according to (\ref{KPZ_quenched})
yields an invariant $\widetilde{\theta_s/d_h} = 0$ for the $\pm J$ model and a
renormalised value $\widetilde{\theta_s/d_h} = -0.0886$ from the $\theta_s/d =
-0.1422$ found above for the Gaussian bonds. Our results of domain-wall energy
computations for $\phi^3$ random graphs of sizes $N_2 = 16^2$, $24^2$, $\ldots$,
$256^2$ are collected in Fig.~\ref{fig:stiffness} together with fits of the form
$\langle |\Delta E|_\mathrm{P/AP}\rangle = E_0 + S_E N_2^{\theta_s/d_h}$. As for the
honeycomb lattice, the Gaussian spin glass shows clear scaling $\langle |\Delta
E|_\mathrm{P/AP}\rangle \rightarrow 0$, with an estimated $\theta_s/d_h =
-0.0684(25)$, different from the ordered lattice case, but also not quite compatible
with the prediction of the quenched approximation (\ref{KPZ_quenched}). The $\pm J$
spin glass {\em as well as\/} the antiferromagnet show saturation behaviour implying
$\theta_s/d_h = 0$ with $E_0 = 0.599(23)$ ($\pm J$) and $E_0 = 0.472(44)$
(antiferromagnet), in agreement with the quenched KPZ prediction. In fact, also the
finite-size approach is very similar between these two cases, with essentially only
the correction amplitude $S_E$ differing, cf.\ Fig.~\ref{fig:stiffness}. This clearly
supports the view that, on the planar random graphs considered here, the
antiferromagnet shows zero-temperature spin-glass behaviour in the universality class
of the $\pm J$ model, in contrast to regular lattices, where the antiferromagnet does
not behave like a spin glass \cite{wannier:50}.

\subsection{Domain-wall fractal dimension}

\begin{figure}[tb]
  \centering
  \includegraphics[clip=true,keepaspectratio=true,width=7cm]{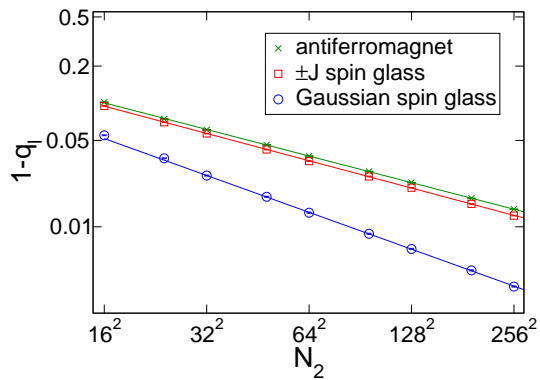}
  \caption
  {Scaling of zero-temperature domain-wall lengths for the Ising antiferromagnet as
    well as the $\pm J$ and Gaussian spin glass on fat $\phi^3$ random graphs. The
    lines show fits of the form (\ref{eq:fractal}) to the data.
    \label{fig:fractal}}
\end{figure}

It has been known for some time that domain walls in spin glasses are fractal curves
\cite{bray:87,kawashima:03a}. Recently, evidence has been presented that in two
dimensions they are in fact in agreement with the excursions known as
``stochastic Loewner evolution'' (SLE), which are closely related with unitary
conformal field theories \cite{amoruso:06a,bernard:06}. It is interesting to measure
the fractal behaviour of domain walls on random graphs which are fractals themselves.
Considering the link overlap $q_l = \frac{1}{N_1} \sum_{\langle i,j\rangle}
s_i^{(P)}s_j^{(P)} s_i^{(AP)}s_j^{(AP)}$ for P/AP boundaries, it is clear that for
Ising spins bonds crossed by a domain wall contribute $-1$ to $q_l$, whereas all
other bonds contribute $+1$ ($N_1$ denotes the number of links of the graph). Hence
one expects the scaling
\begin{equation}
  \langle 1-q_l\rangle \sim \frac{N_2^{d_s/d_h}}{N_1} \sim N_2^{-(1-d_s/d_h)},
  \label{eq:fractal}
\end{equation}
where $d_s$ denotes the fractal dimension of the domain wall (note that $N_1 =
3N_2/2$ from the Euler formula). For the Gaussian spin glass on the honeycomb
lattice, this approach yields $d_s = 1.2725(33)$, in good agreement with the accepted
value for the square lattice \cite{kawashima:03a,amoruso:06a,melchert:07}.  (Here,
the smallest lattice sizes with $L \le 32$ have been omitted from the fit to account
for scaling corrections not explicitly taken into account.) For the bimodal
distribution, the comparison of P and AP ground states does not define a unique
domain wall due to the high degree of accidental degeneracy in the ground state, such
that the above description randomly (but not necessarily without bias) captures one
of these walls and a fit to the form (\ref{eq:fractal}) yields $d_h = 1.283(11)$.
Note that hence there is some uncertainty as to how to define and measure the fractal
dimension properly in this case, and consideration of the backbone of ground states
\cite{roma:07}, of domain walls of extremal length \cite{melchert:07} or of the
residual ground-state entropy \cite{fisch:06a} lead to different estimates for the
square-lattice system.  Noting that $d_s/d_h = 1-\Delta_\mathrm{DW}$, the quenched
KPZ approximation (\ref{KPZ_quenched}) predicts $\widetilde{d_s/d_h} = 0.4666$ for
the Gaussian spin glass coupled to the $\phi^3$ random graphs. The numerical results
shown in Fig.~\ref{fig:fractal} reveal clear scaling for this case with a resulting
estimate $d_s/d_h = 0.5042(24)$, again different from the regular lattice value
$d_s/d = 0.63248(88)$ as well as from the quenched KPZ prediction. The corresponding
estimates of $d_s/d_h = 0.6330(10)$ and $d_s/d_h = 0.6425(10)$ for the $\pm J$ and
antiferromagnetic models due to degeneracies again refer to some random choice of
domain walls. More importantly, however, the scaling approaches for both cases are
again almost identical, cf.\ Fig.~\ref{fig:fractal}, supporting the view of a common
universality class for both models.

\section{Conclusions\label{sec:concl}}

We have shown that the strongly frustrating influence of odd-length loops present in
random graphs can trigger very interesting and characteristic effects in coupled
antiferromagnets, with the observed behaviour often being very different from that
expected for ferromagnets. Only for the special case of bipartite random lattices
without odd-length loops such as dynamical quadrangulations, does a Mattis
transformation identically relate the antiferromagnet to the ferromagnet. The
annealed random-graph average considered in quantum gravity allows the lattice to
dynamically alleviate frustration, leading, for instance, to the emergence of a
N\'eel ordered phase for the $\phi^3$ and $\phi^4$ antiferromagnet, apparently
accompanied by a Kosterlitz-Thouless phase transition (although, of course, it is not
proven that this transition survives in the thermodynamic limit). Quenching the
graphs, such dynamical adaptation is precluded, moving the phase transition to zero
temperature, where instead a spin-glass phase appears. From an exact defect-wall
calculation, it appears that this spin glass is in the same universality class as the
$\pm J$ model on the same lattice. The wealth of these observed effects crucially
depends on the locality (or finite dimensionality) of the considered surfaces, thus
avoiding the possibly less interesting mean-field behaviour induced by more generic
random graphs.

\section*{Acknowledgements}

This work was partially supported by the EC RTN-Network `ENRAGE': {\em Random
  Geometry and Random Matrices: From Quantum Gravity to Econophysics\/} under grant
No.~MRTN-CT-2004-005616. M.W.\ acknowledges support by the EC ``Marie Curie
Individual Intra-European Fellowships'' programme under contract No.\
MEIF-CT-2004-501422.

%\vspace*{-0.5cm}

\vfill

%\bibliography{general}

\end{document}